\documentclass[a4paper]{ESASPCS13Style}
\usepackage{psfig}

\begin{document}

\title{Loop length and magnetic field estimates from oscillations detected
  during an X-ray flare on \object{AT Mic}}

\author{U. Mitra-Kraev \and L.\,K. Harra} \institute{Mullard Space
  Science Laboratory, University College London, Holmbury St.~Mary,
  Dorking, Surrey RH5 6NT, UK} 

\maketitle 

\begin{abstract}

We analyse oscillations observed in the X-ray light curve of the
late-type star \object{AT Mic}. The oscillations occurred during flare
maximum. We interpret these oscillations as density perturbations in
the flare loop. Applying various models derived for the Sun, the loop
length and the magnetic field of the flare can be estimated. We find a
period of 740~s, and that the models give similar results (within a factor of
2) for the loop length ($\sim 5.4\ 10^{10}~{\rm cm}$) and the magnetic
field ($\sim 100~{\rm G}$). 
For the first time, an oscillation of a stellar X-ray flare has
been observed and results thus obtained for otherwise unobservable
physical parameters.  

\end{abstract}

\section{Introduction} \label{intro}
  
Oscillations in the solar corona have been observed for many
years. Wavelength regimes ranging from hard X-ray right down to radio
have been investigated to search for evidence of waves. Periods have
been found ranging from 0.02 to 1000 s. Table 1 in
\cite*{aschwanden1999b} provides an excellent summary of the different
periods that have been found, and an explanation for their
existence. Most of these waves have been explained by MHD oscillations
in coronal loops. \cite*{roberts2000} has provided an
excellent review of waves and oscillations in the corona.  

Many of the observations of waves have been determined from variations
in intensity brightness.  
Recently, however, a huge step forward has been achieved in solar
coronal physics due to the high spatial resolution available with the
Transition Region and Coronal Explorer (TRACE). The first spatial
displacement oscillations have been observed in
coronal loops (\cite{aschwanden1999b}). It was suggested that these
oscillations were triggered by a disturbance from the core flare
site. Various MHD waves were investigated and it was found that a fast
kink mode wave provides the best agreement with the observed period of
280$\pm$30s.  

One of the most exciting aspects of observing waves in this fashion is
that it potentially provides us with the capability of determining the
magnetic field in the corona. It is notoriously difficult to measure
the magnetic field in the corona. Techniques using the near-infrared
emission lines have been successful, but have poor spatial
resolution. Most frequently, indirect methods are used such as the
extrapolations of the coronal magnetic field from the photospheric
magnetic field that can be measured using Zeeman
splitting. \cite*{nakariakov2001} made use of the flare-related
spatial oscillations 
sometimes observed when a flare occurs, to determine the magnetic
field. They assume that the oscillation is due to a global standing
kink wave, and hence the magnetic field is related to the period of
the oscillation, the density of the loop, and the length of the
loop. In the case of \cite*{nakariakov2001} they found the
magnetic field to be 13$\pm$9G. The errors on this are large
due to errors on the determination of density, loop length and
period. \cite*{schrijver2000} suggest that the
oscillations are 
due to a sudden displacement of the magnetic field at the surface which
causes an oscillatory relaxation of the field. Recent work by
\cite*{nakariakov2004} suggests that the oscillations are second
standing harmonics of acoustic waves tied to the loop length. 
The model gives a relationship between the oscillation period, the loop
length and the average plasma temperature, all
of which can be independently observed on the Sun.  

Waves are observed across the electromagnetic spectrum on the
Sun. However, observations on other stars are rare. One reason for
this is that on the Sun, we have the spatial resolution to zoom in on
small regions. For example, the transverse TRACE loops that have been
observed are away from the main flare site, and hence have lower
emission measure. Analysing an X-ray light-curve, for example, would
be unlikely to provide evidence of waves if the emission was averaged
over the whole Sun. Oscillations have been observed for optical
stellar flares (e.g., \cite{mathioudakis2003} and \cite{mullan1992}). 
\cite*{mathioudakis2003} found a period of 220~s in the decay 
phase of a white-light flare on the RS CVn binary \object{II Peg}. Using this
they determine a magnetic field of 1200~G. \cite*{mullan1992} again
found oscillations in X-ray 
active red  
dwarfs. They compared their observations to the possibility that they
were observing p-modes. On the Sun p-modes give rise to 5-minute
oscillations on the surface. They concluded that p-modes could not
produce the amplitudes they observed and that it must be coronal
oscillations, which have, until now, not been observed.

In this work we observe for the first time an oscillation in the
corona of \object{AT Mic}. We measure the periodicity, and hence determine
the magnetic field.  
This is the first observation of an X-ray oscillation during
a stellar flare.

\section{Target}
\object{AT Mic} (\object{GJ 799A/B}) is an M-type binary dwarf, with
both stars of the 
same spectral type (dM4.5e+dM4.5e).  
Both components of the binary flare frequently. 
The radius of \object{AT Mic} given by \cite*{lim1987} is $2.6~10^{10}~{\rm
  cm}$, using a stellar distance of 8.14~pc (\cite{gliese1991}).
Correcting for the newer value for the distance from HIPPARCOS
($10.2\pm0.5~{\rm pc}$, \cite{perryman1997}), using that the stellar
radius is proportional to the stellar distance for a given luminosity
and spectral class, we obtain a stellar
radius of $r_\star = 3.3~10^{10}~{\rm cm} = 0.47~r_{\sun}$. The mass
of \object{AT Mic} given by \cite*{lim1987} is $m_\star = 0.4~m_{\sun}$. 

\section{Observation}
For our analysis we used the {\sl XMM-Newton} observations of \object{AT Mic}
on 16 October 2000 during revolution 156. 
\cite*{raassen2003} have analysed this data spectroscopically,
obtaining elemental abundances, temperatures, densities and emission
measures, while a comparative flare analysis between X-ray and
simultaneously observed ultraviolet emissions can be found in
\cite*{mitra-kraev2004}. 
Here, we solely used the 0.2--12~keV X-ray data from the pn-European Photon
Imaging Camera (EPIC-pn). 

\begin{figure*}
\begin{center}
\psfig{figure=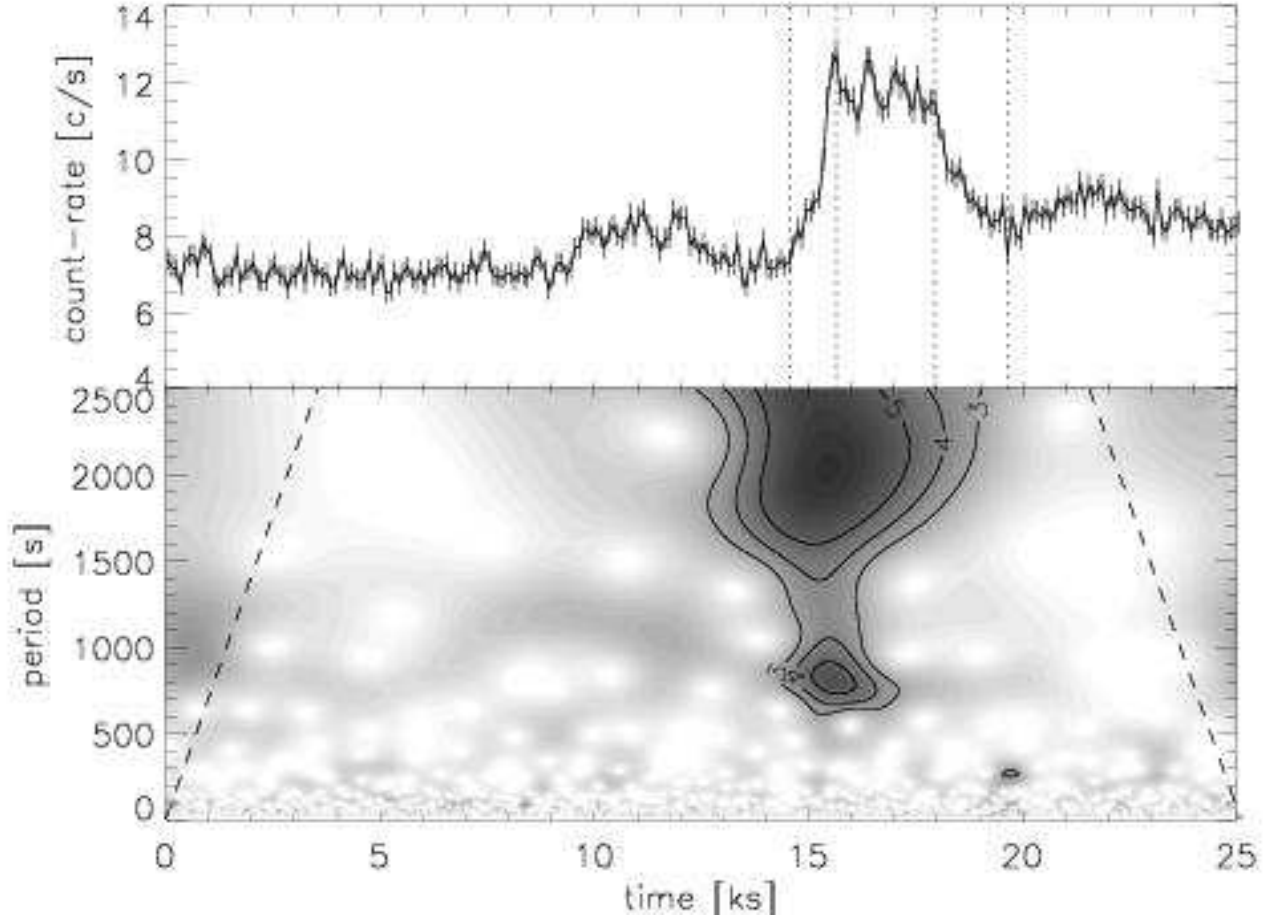}
\caption{The light curve and wavelet coefficients of \object{AT Mic}. The
  observation started on 2000-10-16 at 00:42:00~h and lasted for
  25'100~s ($\sim$ 7~h). The upper panel shows the observed light
  curve, binned up to 100~s. The vertical dashed lines indicate the
  start and end of the rise and decay phase of the flare. The lower
  panel shows the absolute values of a section of the corresponding
  wavelet coefficients (see main text) divided by their standard 
  deviation. The 3, 4 and 5 $\sigma$ significance contour lines are
  drawn. The two dashed lines mark the border of the cone of
  influence. A local maximum is clearly seen with a period of around
  740~s during the flare peak. 
  \label{figure}} 
\end{center}
\end{figure*}
Figure \ref{figure} shows the \object{AT Mic} light curve. 
The observation started at 00:42:00 and lasted for 25.1~ks ($\sim$
7~h).
There is one large flare, starting $\sim$ 4~h into the observation,
increasing the count-rate from flare onset to flare peak by a factor
of 1.7, and lasting for 1~h~25~min.
It shows a steep rise (rise time $\tau_r = 1100~{\rm s}$) and decay
(decay time $\tau_d = 1700~{\rm s}$). There is an extended peak to
this flare, which shows clear oscillatory behaviour. The amplitude of
the oscillation is around 16\%.  
Applying multi-temperature fitting, \cite*{raassen2003}
obtain a mean flare temperature $T = 24\pm 4~{\rm MK}$ and a quiescent
temperature $T_q = 13\pm 1~{\rm MK}$, and from the \ion{O}{vii} line
ratio a flare and quiescent electron density of $n =
4^{+5}_{-3}~10^{10}~{\rm cm}^{-3}$ and $n_q = 1.9\pm 1.5~10^{10}~{\rm
  cm}^{-3}$, respectively. The total flare and quiescent emission
measures are $EM = (19.5\pm0.8)10^{51}~{\rm cm^{-3}}$ and $EM_q =
(12.2\pm0.5)10^{51}~{\rm cm^{-3}}$.

One of the reasons why we concentrated on this
observation is that following the rise phase of the flare there was a
period of approximately 40 minutes when the light curve stayed at a
high intensity level and showed strong oscillations. The oscillatory
activity can be clearly seen by eye (see Fig.~\ref{figure}), and shows
behaviour suggestive of damping. 
This character is much different from solar flares where the flare
reaches a peak rapidly, followed by a slow decay. 
For example, \cite*{svestka} shows solar flare light curves
and a flat top is not seen. 
An alternative to oscillations for this flare is that repeated and rapid flaring is occurring. We do not consider this option in this paper, and assume for our analysis that we are observing a coronal oscillation.
The purpose of our analysis is to determine through wavelet analysis
the period and amplitude of the oscillation. 
We determine the magnetic field and loop length of the coronal loop
assuming that the oscillation is due to an acoustic wave. 
As a validity check the value of loop length was compared to the
value determined from a radiative cooling model.  

\section{Results}

\subsection{Oscillation during flare maximum} \label{wt}

To investigate periodicities during the observation, we applied a
continuous wavelet transformation to the 10s-bin\-ned-data, using a Morlet
wavelet (see, e.g., \cite{torrence1998}). 
The wavelet coefficients with the lowest periods are displayed in
Fig.~\ref{figure}. 
The flare oscillation is picked up with a 5$\sigma$ significance level.
The local maximum with the shortest period and a significance of more than
5$\sigma$ is simultaneous with the flare top and identifies the
oscillation, which, at the local maximum, has a period of $P=740~{\rm
  s}$ . The period interval 
where the wavelet coefficients have a significance of 5$\sigma$ or
more, is [730,920]~s.

\subsection{Loop length from oscillation} \label{oscill}
Interpreting these oscillations as the second spatial harmonics of an
acoustic wave within the flare (\cite{nakariakov2004}), we can then
determine the flare loop length from
\begin{equation}
L\ [{\rm Mm}] \approx \frac{1}{6.7} \cdot P\ [{\rm s}] \cdot \sqrt{T\
  [{\rm MK}]}. \label{ll}
\end{equation}
Inserting the above obtained period $P = 740~{\rm s}$ and a flare
temperature $T = 24~{\rm MK}$, we obtain a loop length $L =
5.4~10^{10}~{\rm cm}$. Note that the temperature carries a $1\sigma$
error of $\sim 16\%$, therefore the error of $L$ is at least 8\%
($\Delta L > 0.5~10^{10}~{\rm cm}$). It is not straightforward to
calculate the corresponding $1\sigma$ error of $P$. 
With these numbers, the inferred speed of sound is $c_s = L/P \approx
730~{\rm km/s}$.

\subsection{Loop length from radiative cooling times} \label{radcool}
The loop length can also (and independently of any oscillation) be
estimated from rising and cooling times obtained from the temporal
shape of the flare, applying a flare heating/cooling model (see, e.g.,
\cite{cargill1995}). We follow
the approach by \cite*{hawley1995} who investigated a flare on AD Leo
observed in the extreme ultraviolet. The shape of this flare is very
similar to our flare on \object{AT Mic}, but roughly 10 times larger (in
duration as well as in the increase of the count rate). It also shows
a flat top with a possible oscillation. The flare loop energy equation for
the spatial average is given by
\begin{equation}
\frac{3}{2}\ \dot{p} = Q - R,
\end{equation}
with $Q$ the volumetric flare heating rate, $R$ the optically thin
cooling rate and $\dot{p}$ the time rate change of the loop
pressure. During the rise phase, strong evaporative
heating is dominant ($Q\gg R$), while the decay phase is dominated by
radiative cooling and strong condensation ($R \gg Q$). 
At the loop top, there is an
equilibrium ($R = Q$). The loop length can be derived as
\begin{equation}
L = \frac{1500}{\left( 1-x_d^{1.58} \right)^{4/7}} \cdot \tau_d^{4/7}
  \cdot \tau_r^{3/7} \cdot T^{1/2},
\end{equation}
where $\tau_r$ is the rise time, $\tau_d$ the flare decay time
(indicated in Fig.~\ref{figure} with the vertical dotted lines), $T$
the apex flare temperature and $x_d^2=c_d/c_{max}$, with $c_{max}$
the peak count rate and $c_d$ the count rate at the end of the
flare. Inserting these values, the loop length becomes
$L=2.3~10^{10}~{\rm cm}$, which is about a factor of 2 smaller that
the loop length inferred from the oscillation.

\subsection{Determination of the Magnetic field}
Interpreting the oscillations in terms of global standing kink waves,
\cite*{nakariakov2001} derive a relation for the magnetic field
\begin{equation}
B = \sqrt{8\pi}\ \frac{L}{P}\ \sqrt{\rho\left( 1 +
    \frac{\rho_{ext}}{\rho}\right)}, \label{mag}
\end{equation}
with $\rho$ the mass density inside, and $\rho_{ext}$ the mass density
outside the loop. Note that there is a correction factor of 0.64 in
the equation as presented by \cite*{nakariakov2001}
(\cite{mathioudakis2003}). 
We can simplify Eq.~(\ref{mag}) by inserting Eq.~(\ref{ll}) and using
$\rho = m_p\cdot n$ and $\rho_{ext} = m_p\cdot n_q$, with $m_p$ the
proton mass. Then, the magnetic field is given by
\begin{equation}
B = \frac{\sqrt{8\pi}}{6.7}\cdot 10^5\cdot \sqrt{T m_p n\left(
    1+\frac{n_q}{n}\right)} . 
\end{equation}
Inserting the temperature and densities from the spectroscopy results,
we obtain a magnetic field of $B = 115\pm56~{\rm G}$. Note that the large
error is resulting mainly from the density uncertainty. 
Using the loop length from the radiative cooling approach
(Sect.~\ref{radcool}) and Eq.~\ref{mag}, the magnetic field is
$B\approx 50~{\rm G}$.

\subsection{Pressure balance} \label{pressbal}
To maintain stable flare loops, the gas pressure of the evaporated
plasma must be smaller than the magnetic pressure
\begin{equation}
2nkT \leq \frac{B^2}{8\pi}.
\end{equation}
Knowing the flare density and temperature, we get a lower limit for
the magnetic field $B>80\pm 60~{\rm G}$. Again, there is a large error
because of the large uncertainty for the density. \cite*{shibata2002}
assume pressure balance and give equations for $B(EM,n_q,T)$ and
$L(EM,n_q,T)$ (their Eqs~7a,b). Using these relations, we obtain a
magnetic field of
$B=70\pm40~{\rm G}$ and a loop length of $L=(2.8\pm1.7)\ 10^{10}~{\rm cm}$. 

\section{Discussion and Conclusions} 
We have used three different approaches to determine the loop length
of the \object{AT Mic} flare and two different approaches for the magnetic
field. The loop length derived in Sect.~\ref{oscill} from the
flare oscillation is the largest with $L = 5.4\cdot
10^{10}~{\rm cm}$, while the loop length derived from radiative
cooling times (Sect.~\ref{radcool}) is somewhat smaller by roughly a
factor of 2, $L=2.3~10^{10}~{\rm cm}$. The loop length derived from
pressure balance (Sect.~\ref{pressbal}) is similar to the latter one, 
$L=(2.8\pm1.7)\ 10^{10}~{\rm cm}$. The loop length derived by assuming
that the variations seen are due to oscillations is a factor of 2
different to the loop length determined by more usual
methods. Considering the assumptions that have to be made, this is
good agreement, and gives us confidence that we are, for the
first time, observing a stellar coronal loop oscillating.  
 The magnetic field ranges
from 50~G (radiative cooling) to 115$\pm$56~G (oscillation), with the
value from pressure balance in between with 70$\pm$40~G. Again, these
three values are all consistent.

This was the first time that an oscillation during flare peak was
observed in X-rays in a stellar flare and flare loop length and
magnetic field derived from it. The values are consistent with other
flare models.

\begin{acknowledgements}
We ac\-knowl\-edge fi\-nan\-cial sup\-port from the UK Par\-ti\-cle
Phy\-sics and Astronomy 
Research Council (PPARC). UMK would also like to thank the European
Space Agency (ESA) and the University College London (UCL) Graduate
School for financial assistance to attend the Cool Stars 13 conference. 
\end{acknowledgements}

\appendix


\begin{thebibliography}{}

\bibitem[\protect\astroncite{Aschwanden et~al.}{1999}]{aschwanden1999b}
Aschwanden M.J., Fletcher L., C. J. Schrijver et al.\  1999, ApJ 520, 880

\bibitem[\protect\astroncite{Cargill et~al.}{1995}]{cargill1995}
Cargill P.J., Mariska J.T., Antiochos S.K. 1995, ApJ 439, 1034

\bibitem[\protect\astroncite{Gliese \& Jahreiss}{1991}]{gliese1991}
Gliese W., Jahreiss H. 1991, Preliminary Version of the Third
Catalogue of Nearby Stars, Astron. Rechen-Institut, Heidelberg

\bibitem[\protect\astroncite{Hawley et~al.}{1995}]{hawley1995}
Hawley S.L., Fisher G.H., Simon T. et al.\ 1995, ApJ 453, 464

\bibitem[\protect\astroncite{Lim et~al.}{1987}]{lim1987}
Lim J., Nelson G.J., Vaughan A.E. 1987, Proc. ASA 7, 2

\bibitem[\protect\astroncite{Mathioudakis et~al.}{2003}]{mathioudakis2003}
Mathioudakis M., Seiradakis J.H., Williams D.R. et al.\ 2003, A\&A 403, 1101

\bibitem[\protect\astroncite{Mitra-Kraev et~al.}{2004}]{mitra-kraev2004}
Mitra-Kraev U., Harra L.K., G\"udel M. et al.\ 2004, A\&A, submitted

\bibitem[\protect\astroncite{Mullan et~al.}{1992}]{mullan1992}
Mullan D.J., Herr R.B., Bhattacharyya S. 1992, ApJ 391, 265

\bibitem[\protect\astroncite{Nakariakov \& Ofman}{2001}]{nakariakov2001}
Nakariakov V.M., Ofman L. 2001, A\&A 372, L53

\bibitem[\protect\astroncite{Nakariakov et~al.}{2004}]{nakariakov2004}
Nakariakov V.M., Tsiklauri D., Kelly A. et al.\ 2004, A\&A 414, L25

\bibitem[\protect\astroncite{Perryman et~al.}{1997}]{perryman1997}
Perryman M.A.C., Lindegren L., Kovalesky J. et al.\ 1997, A\&A 323, L49

\bibitem[\protect\astroncite{Raassen et~al.}{2003}]{raassen2003}
Raassen A.J.J., Mewe R., Audard M. et al.\ 2003, A\&A 411, 509

\bibitem[\protect\astroncite{Roberts}{2000}]{roberts2000}
Roberts B. 2000, Sol. Phys. 193, 139

\bibitem[\protect\astroncite{Schrijver \& Brown}{2000}]{schrijver2000}
Schrijver C.J., Brown D.S. 2000, ApJ 537, L69

\bibitem[\protect\astroncite{Shibata \& Yokoyama}{2002}]{shibata2002}
Shibata K., Yokoyama T. 2002, ApJ 577, 422

\bibitem[\protect\astroncite{Svestka}{1989}]{svestka}
Svestka, Z., 1989, Sol. Phys. 121, 399.

\bibitem[\protect\astroncite{Torrence \& Compo}{1998}]{torrence1998}
Torrence C., Compo G.P. 1998, Bull. Amer. Meteor. Soc. 79, 61


\end{thebibliography}
\end{document}